# Intelligent learning environments within Blended Learning for Ensuring Effective C Programming Course

Utku Kose[1] and Omer Deperlioglu[2]

[1]Afyon Kocatepe University, Distance Education Vocational School,
ANS Campus, 03200, Afyonkarahisar, Turkey
`utkukose@aku.edu.tr`
[2]Afyon Kocatepe University, Engineering Faculty,
ANS Campus, 03200, Afyonkarahisar, Turkey
`odeper@aku.edu.tr`

## ABSTRACT

*This paper describes a blended learning implementation and experience supported with intelligent learning environments included in a learning management system (LMS) called @KU-UZEM. The blended learning model is realized as a combination of face to face education and e-learning. The intelligent learning environments consist of two applications named CTutor, ITest. In addition to standard e-learning tools, students can use CTutor to resolve C programming exercises. CTutor is a problem-solving environment, which diagnoses students' knowledge level but also gives feedbacks and tips to help them to understand the course subject, overcome their misconceptions and reinforce learnt concepts. ITest provides an assessment environment in which students can take quizzes that were prepared according to their learning levels. The realized model was used for two terms in the "C Programming" course given at Afyon Kocatepe University. A survey was conducted at the end of the course to find out to what extent the students were accepting the blended learning model supported with @KU-UZEM and to discover students' attitude towards intelligent learning environments. Additionally, an experiment formed with an experimental group who took an active part in the realized model and a control group who only took the face to face education was performed during the first term of the course. According to the results, students were satisfied with intelligent learning environments and the realized learning model. Furthermore, the use of intelligent learning environments improved the students' knowledge about C programming.*

## KEYWORDS

*Intelligent learning environments, knowledge-based systems, blended learning, learning reinforcement, assessment for learning*

## 1. INTRODUCTION

Intelligent learning environments are some kind of software solutions, which are designed to facilitate effective learning through the exploration of complex problems. The main objective of intelligent learning environments is to improve students' learning process by providing more advanced educational techniques. In this aim, intelligent learning environments support both learning through reflective discovery as well as monitored goal-directed interaction [1]. Thus, intelligent learning environments are concerned with students developing both general and domain specific thinking and problem solving skills [1, 2]. Nowadays, intelligent learning environments are widely used in modern educational approaches and systems. Because they promote more effective learning than the traditional (face to face) instruction [3].

In this paper, a blended learning implementation and experience supported with two intelligent learning environments included in a learning management system (LMS) called @KU-UZEM is described. The study is based on a model, which was used for two terms in the "C Programming" course given at Afyon Kocatepe University in Turkey. With the realized model, a combination of face to face education and e-learning is provided to perform educational activities. E-learning is realized by using University's Learning Management System (LMS) named @KU-UZEM. @KU-UZEM consists of many e-learning tools like communication modules, lesson modules, security applications, counseling services, management modules etc... In addition to these tools, students can use two intelligent learning environments in e-learning activities. Intelligent learning environments provided by @KU-UZEM are named as CTutor and ITest respectively. CTutor enables students to resolve C programming exercises. CTutor is a problem-solving environment, which diagnoses students' knowledge level but also gives feedbacks and tips to help them to understand the course subject, overcome their misconceptions and reinforce learnt concepts. On the other hand, ITest provides an assessment environment in which students can take quizzes that were prepared according to each student's learning level.

The main objective of this study is to evaluate the blended learning model and the intelligent learning environments that enable students and teachers to perform more effective and efficient educational activities. The course subject handled in this study is an essential course given for computer programming students and it includes important practical and theoretical topics that students must know to learn C programming. At this point, it is an important approach to use intelligent learning environments that support students' practical learning activities in such courses and enable them to improve their academic achievements with more effective techniques. This study also aims to raise the interest level in intelligent learning environments, blended learning and e-learning among computer programming students and teachers.

The rest of the paper is organized as follows: Section 2 describes the pedagogical foundations of the study. The developed intelligent learning environments (CTutor, ITest) are tackled in detail in Section 3. Section 4 is devoted to design and development of the blended "C Programming" course, which was implemented within the study. Section 5 describes the survey, which was conducted at the end of the course to find out to what extent the students were accepting the blended learning model supported with @KU-UZEM and to discover students' attitude towards intelligent learning environments. In addition to the survey, the experiment, which was formed with an experimental group who took an active part in the realized model and a control group who only took the face to face education, is also described in this section. Finally, Section 6 outlines the conclusions that have been obtained with this study and gives some information about future works.

## 2. PEDAGOGICAL FOUNDATIONS

As mentioned before, the main objective of intelligent learning environments is to improve students' learning process by providing more advanced educational techniques. So, working principles of intelligent learning environments are based on different pedagogical approaches and theories that have been developed in the education field. In this sense, feedback is some of the most important approaches that intelligent learning environments are mostly built on.

Feedback is an essential construct for many theories of both learning and instruction, and an understanding of the conditions for effective feedback should facilitate both theoretical development and instructional practice. The most principle function of feedback is to ensure more effective instructional design for learning activities and improve students' performance [4]. In the literature, feedback has been regarded as one of the most critical sources of information to assist students in restructuring their own knowledge and support their metacognitive processes [5 - 7]. In this aim, feedback approach helps students in a variety of ways. It is important that it helps students to evaluate their own performance and enable them to see what they do well and what they need to improve. Furthermore, a feedback system encourages learners to try new skills: they can challenge themselves, experiment with new skills, and receive guidance that helps them develop mastery before being graded [8]. Eventually, all of these functions make feedback an indispensible factor for especially computer-assisted learning systems. Naturally, intelligent learning environments are one of these systems that use feedback approach for better learning experiences.

In the intelligent learning environments, three approaches of feedback can be used to provide information to students. These approaches are called as "knowledge of response", which simply informs the student about a correct or incorrect response; "knowledge of correct response", which informs the student about what should be the correct response; and "elaborated feedback", which provides an explanation for why the student's response is correct or incorrect, or allows the student to review material relevant to the attributes of a correct response [9]. In addition to the standard form of feedback, Sales has also suggested the "adapted feedback", in which special attention was put into the role of feedback and an effort was made to customize the feedback along one or more dimensions, to compensate for the weakness of the standard feedback [10].

Today, feedback approach is widely used in many educational activities to provide more effective learning processes for students. Especially, it is a necessary approach for the courses, which are based on technical and practical subjects. It is clear that students can learn abstract concepts and perform practical exercises better when they receive immediate feedbacks. Otherwise, both teachers and students come across many difficulties during the education process.

Originating point of this study is difficulties in teaching computer programming. These difficulties are mostly based on intelligibility levels of course subjects. Intelligibility levels can be adjusted to be suitable for students by choosing effective learning approaches and using feedbacks. Intelligent learning environments are one of the most effective learning tools that can be used to reduce mentioned difficulties and provide more effective learning experiences for students. Before talking about intelligent learning environments for learning computer programming, it is better to show up difficulties in teaching computer programming.

Students, who learn computer programming for the first time, can find it difficult to understand algorithmic thinking and principles of computer programming. Computer programming is connected with many abstract and theoretical subjects that need more advanced studies to be learned. Especially in programming language courses, students often find it difficult to use their knowledge for designing and developing computer programs. Another important factor that triggers the mentioned problem is complexity of today's programming languages. When computing was first introduced and performed mechanically, programming vocabulary consisted of the sequencing of basic steps. However, the subsequent addition of procedural abstraction and object orientation allowed more abstract conceptualization of computing [11]. Hence, more advanced and complex programming techniques and languages have been developed. Today, educational institutes provide the latest course contents about computer programming. As a result, courses cause some students to become anxious, or even to fear programming. As performance is negatively affected by the anxiety, the impact on academic performance has negative effects on retention rates. Moreover, uncontrolled anxiety levels affect students' ability and cause poor academic progress with high dropout rates [12].

Many different research studies have been made to provide effective solutions for these problems in teaching computer programming. In addition to the theoretical approaches, there exist some intelligent tools that have been designed and developed to teach programming concepts in different programming languages. The most popular ones are the family of ACT tutors [13 - 16], which are intelligent tools developed for learning Pascal, Lisp and Prolog programming languages. Another tool named ViRPlay [17] is centered on interactions in programs written in Java. ViRPlay tries to teach the execution of a program, using role-play simulations in a developed virtual 3D environment. ELM-PE [18] is an intelligent learning environment that was developed as an example-based learning system for Lisp. Finally, COLLECT-UML [19] is an intelligent tool that was developed for learning object oriented programming.

Although these tools provide effective learning environments for students, education process can also be improved by using them in a course, which is planned with different educational approaches and models. At this point, it can be an effective teaching approach to use intelligent learning environments within an effective and advanced learning model.

In the face to face education, teaching and learning activities are performed in the classrooms where teachers contact with students and give feedback for performed activities. Today, face to face education continues to be most common learning strategy. But teachers have to provide instruction to dozens or even hundreds of students and this situation may cause some problems in making educational activities [16, 20]. For instance, students cannot correctly assimilate the lessons being taught. New education methods and techniques have been developed to solve these problems and provide more effective education experiences for both teachers and students. By using blended learning, teachers have a chance to remove problems of the face to face education and provide better educational environments for students. Blended learning is an education model, which contains different types of learning strategies. The objective of this learning model is to provide better teaching-learning experiences with combinations of learning strategies [21 - 23].

By adopting intelligent learning environments, blended learning models can be improved to ensure more effective and efficient education experiences. With these models, students have a chance to perform both face to face and artificial intelligence supported e-learning activities. In this way, it is possible to improve students' academic achievements and develop an educational model in which students enjoy the provided learning activities. These scenarios have given rise to the approach implemented in this study.

## 3. INTELLIGENT LEARNING ENVIRONMENTS FOR C PROGRAMMING

Within this study, two different intelligent learning environments have been designed and developed for using in C programming course given at Afyon Kocatepe University. These intelligent learning environments are provided by @KU-UZEM and named as CTutor and ITest respectively. It is better to explain using features and functions of both CTutor and ITest to understand their roles in the developed learning model.

### 3.1. CTutor

CTutor is an intelligent learning environment in which students can take some exercises by using an easy to use interface and teachers can define new C programming exercises with the provided tools. Programming exercises, which are created with CTutor, can be provided in any course lesson page included in the @KU-UZEM. Thus, students can take specific exercises that are related to an actual course lesson. It is also possible to view lesson texts by clicking on links provided in course lesson pages of @KU-UZEM.

CTutor allows teachers to create new exercises by using a management interface provided in the system. For each exercise, the teacher can define the problem text and develop what would be the correct solution to that problem in the same way as a student would do. Moreover, domain expert knowledge of the CTutor can also be adjusted for specific exercises by using the management interface of the system. All of these operations can be done easily by using the CTutor interface supported with drag and drop feature and simple system controls.

From students' perspective, the CTutor interface is some kind of tool, which assembles different system controls to get the representation of solutions and deliver exercises, feedbacks and information. While using the CTutor, students must apply their knowledge on C programming to develop solutions for given exercises. Related solutions must be given in a special form, which can be parsed and understood easily by the CTutor system. To achieve this, some specific using features and functions were included in the system. For instance, the system interface allows solution C programs to be built by means of a drag and drop feature, which limits the actions that can be done in the CTutor and focuses students' ideas on solving processes instead of writing C source codes. Eventually, this function permits the CTutor to trace all actions performed by students and facilitates the adaptation, which can be provided by the system.

As mentioned before, CTutor employs an easy to use interface, which can be used by students to take and solve given exercises. In this way, the system aims to ensure an effective and fast learning experience for students. After creating exercises for a specific course lesson, the CTutor is automatically imported to the related lesson page provided on @KU-UZEM. Figure 1 shows a screenshot from the CTutor interface for students.

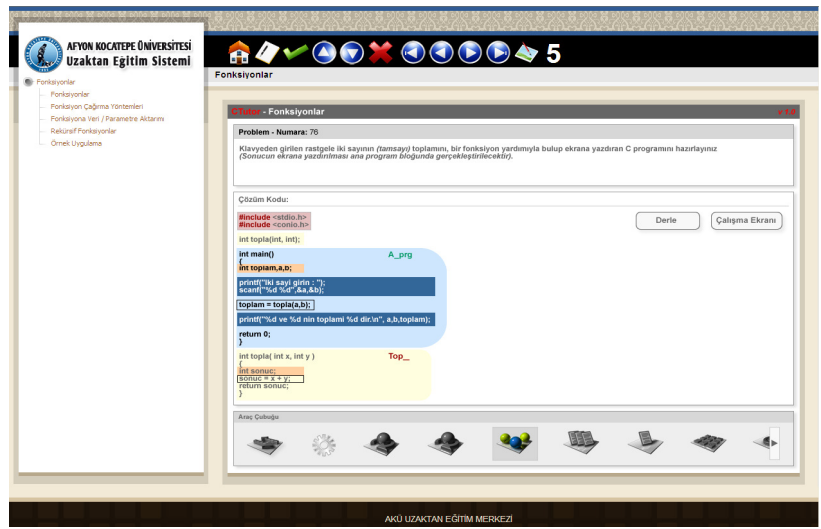

Figure 1. A screenshot from the CTutor interface for students

A typical CTutor interface, which can be viewed by students, consists of three different parts. These parts provide all necessary elements, which enable students to understand problem of the given exercise, develop a possible solution for this exercise and view obtained results. The first part is located on top of the interface and used to show "problem text" of the given exercise. Under this part, the "workspace", where students can develop a solution C code for the given exercise, is located. By using this part, the student can start to create a solution C code or edit the developed one according to the received feedbacks. It is easy to edit any written code (block) by double-clicking on it. The workspace includes two buttons named as "Compile" and "Runtime Screen" respectively. The "Compile" button is used to execute developed solution programs. On the other hand, the "Runtime Screen" button is used to open a new window, where students can view the "runtime" of the solution program after the compiling process. After a successful compiling process, this window is also opened automatically by the CTutor. The last part of the interface, which is also named as "Tool Bar", is located under the workspace. The Tool Bar includes many different elements that can be used to develop a solution C program on the workspace. Students can use the provided elements to define different program parts like declarations, preprocessors, if statements, iteration (loops), functions…etc. Each element can be added to the workspace by dragging the element icon and dropping it into the workspace. When adding an element to the workspace, some additional information such as names, parameters, types…etc. is also requested by the system. CTutor also views different code types in separate "layer" elements. This feature allows teachers and students to understand code structure easily and enables system to evaluate developed programs faster.

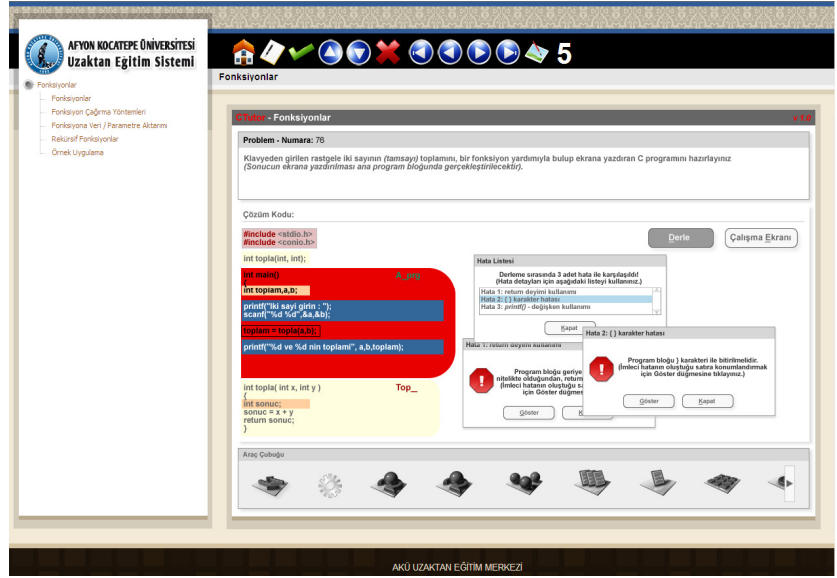

Figure 2. Some feedbacks shown by the CTutor

After developing a solution for the given exercise, the student can start the evaluation process by clicking on the Compile button. After the evaluation process, CTutor gives feedbacks about errors that were made within the solution program (Figure 2). Evaluation mechanism of the CTutor is based on a domain prepared according to the expert knowledge. At this point, the domain of C programming is very complex. There is no fixed sequence of actions that will enable users to get the solution, nor is there only one solution for a given exercise. Indeed, there are an infinite number of C code combinations which will lead the user to a valid solution. So, such systems use a student modeling called "Constraint Based Modeling" to handle the wide space of solutions.

Constraint Based Modeling is based on Ohlsson's theory of learning from errors [24]. This technique proposes that the student can learn from the feedback generated as the result of an error. According to this technique, the domain includes some basic principles, which should be supported by all solutions for a given problem [16]. A constraint based model represents knowledge about a domain as a set of "constraints" on correct solutions. The constraints select, out of the universe of all possible solutions, the set of correct solutions. Eventually, they partition the universe of possible solutions into the correct and the incorrect ones.

A formal notation has been introduced by Ohlsson and Rees to be used for constraints within models [25]. The unit of knowledge is called as a state constraint and each state constraint is used as an ordered pair of <Cr, Cs>, where Cr is the relevance condition and Cs is the satisfaction condition. Cr identifies the class of problem states for which the constraint is relevant whereas Cs identifies the class of (relevant) states in which the constraint is satisfied. Each member of the pair can be thought of as a set of features or properties of a problem state. At the same time, constraints are encoded by rules of the form: IF Cr is satisfied, THEN Cs should also be satisfied; otherwise a principle is being violated. Briefly, the domain model consists of a set of rules, which represent general principles that must not be broken [14]. In the domain model of the CTutor, some examples can be the followings:

- Cr = "a problem requirement is to apply a function to a range of numbers" and Cs = "it must be the case that the solution program contains a loop"

- Cr = "exist an assignation element" and Cs = "it must be the case that there is a valid expression on the right-hand side of the element"

- Cr = "exist an assignation element" and Cs = "data types associated on both sides of the assignation must be equal"

Domain model of the CTutor has been developed by using CLIPS inference rules. The antecedent of any rule is formed by a combination of Cr and Cs conditions whereas the consequent is in charge of throwing an error with the information related to a violated rule. During the evaluation process, errors are captured by the CTutor system and some feedbacks are shown to students via interface. The domain is stored in a knowledge base, which can be managed easily by teachers. As mentioned before, teachers can adjust the knowledge base by using the management interface of the CTutor system. All rules stored in the system have been prepared in collaboration with experts. A total amount of 438 rules have been identified and grouped into eight different categories. Table 1 shows the rule categories of the developed domain model.

Table 1. Rule categories of the developed domain model

| Category Name | Category Description | Rules |
| --- | --- | --- |
| Solution methods | For using rules of some well-known solution methods. | 54 |
| Missing references | For errors that may take place as consequence of code elimination. | 80 |
| Pointer | For errors that may appear while using pointers. | 23 |
| Memory | For errors and using principles about dynamic memory operations. | 12 |
| File | For errors that may appear within file operations (text and binary). | 44 |
| Functions | For using principles of different function types, parameters…etc. | 71 |
| Data types | For incoherencies among the data types of some expressions. | 52 |
| Syntax | For syntactic errors and infraction of grammar rules. | 102 |

As mentioned before, CTutor also uses "layer" elements to determine different code types of any developed C program. Layers used by the CTutor system are defined by using CLIPS templates, which also define indirectly the corresponding code types. Every code type, which is

International Journal of Artificial Intelligence & Applications (IJAIA), Vol.3, No.1, January 2012

widely used on the workspace, has a special template associated. A template has a set of fields that determine the necessary content for every code type. Some of these fields are used to define the characteristics of the template or code type and ensure connection with other elements included in the workspace.

In CTutor, two different types of layers are used to determine C code types:

- Basic layers, which are associated with some standard C codes like basic data types (Integer, Float, Char…etc.) and some standard functions (Printf(), Scanf()…etc.).

- Advanced layers, which are often associated with some special code blocks (loops, if…else, switch…etc.) or algorithms (sorting algorithms, searching algorithms…etc.). These layers are used in the workspace when there is need for some certain program structures and special solution methods to solve the given exercise problem.

Students can gain learning scores (changing between 0 and 100) according to their learning process in the CTutor. Learning scores can be gained according to some criterions like exercise completing time and number of received feedbacks. Limit values for these criterions are defined by the teacher while preparing a CTutor exercise. Students' learning scores are stored in another database included in a module provided by the @KU-UZEM. This module is named as "Performance Tracing Module" and it is used by teachers to store information about students' activities on the @KU-UZEM. Thus, learning scores can be examined and evaluated by teachers on the @KU-UZEM interface. Learning scores are also used by ITest to calculate each student's learning level.

### 3.2. ITest

ITest is an assessment environment that students can use to take special quizzes during the e-learning process. For each quiz, ITest system chooses the most appropriate questions according to each student's learning level. ITest quizzes are provided to students via course lesson pages included in the @KU-UZEM. Teachers can arrange one or more quizzes for students and provide them in any course lesson page according to the e-learning plan of the course. Thus, students have a chance to check their own knowledge and learning level about the given course lessons.

ITest consists of four different modules that determine working mechanism of the whole application. These modules are named as "Quiz Interface", "Question Bank – Question Knowledge Base", "Student Model Base" and "Question Chooser" respectively. It is better to explain these modules to understand working mechanism of the ITest.

"Quiz Interface" module is the page, where students can view the whole quiz questions and answer them by using provided controls. This module is a simple interface that enables students to navigate through the listed questions, read questions and select possible answers easily. After starting a quiz from any course lesson page, related questions are chosen by the "Question Chooser" module and placed automatically into the Quiz Interface module. Quiz Interface module works as integrated to the @KU-UZEM interface. Figure 3 shows a screenshot from a quiz session.

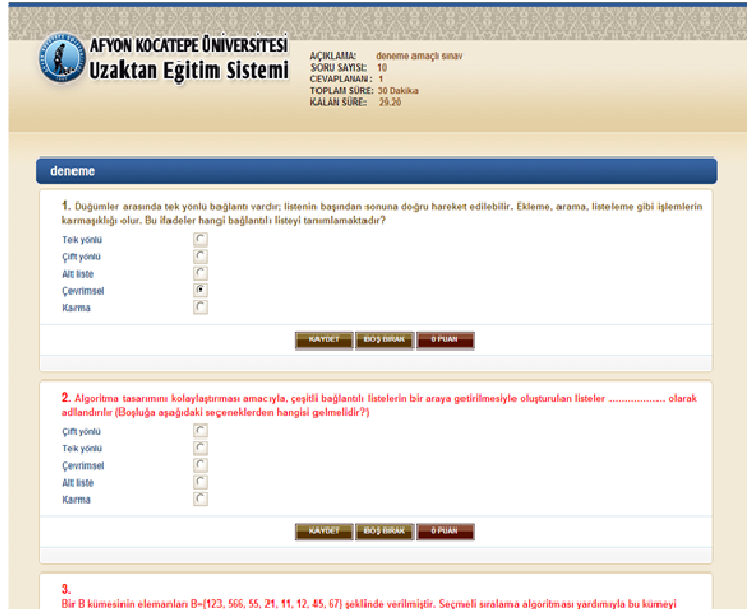

Figure 3. A screenshot from a quiz session

"Question Bank" module is used by teachers to prepare and store quiz questions for future uses. This module can also be called as "Question Knowledge Base". After starting a quiz, questions are automatically chosen from the knowledge base included in this module. Teacher can use a management interface to add new questions to the base or edit the stored ones. It is important that there must be enough questions stored in the base to provide different quizzes for each student. Each quiz question must also be stored under one of pre-defined course lesson titles. A typical ITest question is a multi-choice test item with five possible answers. During the preparation process, teachers can use a rich text editor to prepare questions supported with different text styles and multimedia elements. It is also important that teachers can also define a difficulty level, choice priority and answering time for each question. Questions are classified under five different difficulty levels (1: very easy, 2: easy, 3: normal, 4: hard and 5: very hard) and "Question Chooser" considers the level of each question to determine appropriate questions for any student. Choice priority is a value, which is used by the ITest to choose the most appropriate questions for each student. This value is automatically increased or decreased if the question was used in any quiz. On the other hand, question answering time is used to calculate total answering time of prepared quizzes.

The third module of the ITest is "Student Model Base". This module includes another knowledge base, which stores some necessary information to define each student's quiz performance and learning level. For each student, the base includes information about past quiz scores, number of correct and incorrect answers for each completed quiz and number of completed CTutor exercises. Some of information stored in the Student Model Base is used by the "Question Chooser" module to calculate each student's learning level.

"Question Chooser" module is the most important part of the ITest system. Before viewing the quiz interface for the student, this module calculates the student's learning level and chooses the questions that will be provided. Each student's learning level is calculated by using three different factors: number of views for the course lesson pages, which are related to the provided quiz, past quiz scores and past CTutor learning scores. Page viewing statistics is stored in the Performance Tracing Module of the @KU-UZEM. In this module, teachers can examine and evaluate these statistics and transform them to the values changing from 0 to 100. In other words, teachers give page viewing scores for each student. Thus, ITutor can use the page viewing statistics as page viewing scores to calculate the student's learning level. Question Chooser calculates the student's learning level by summing 10 percent of his / her page viewing score, 40 percent of his / her average past quiz score and 50 percent of his / her average past CTutor learning score. Later, obtained value is used to choose appropriate questions from the Question Bank – Question Knowledge Base" module. It is important that Question Chooser considers the difficulty level and choice priority of each question during this process.

As it is understood, all modules explained above determine the working principles of the ITest. Figure 4 illustrates a schema showing the working mechanism of the ITest briefly.

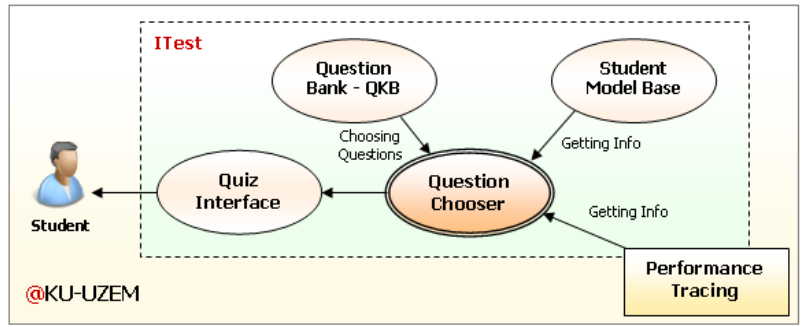

Figure 4. Working mechanism of the ITest

At the end of any quiz, the student's total score is automatically calculated by the ITest and stored to the related knowledge base. ITest system allows teachers to view each student's quiz scores by using the @KU-UZEM interface. After the review process, quiz results can also be announced to students via @KU-UZEM.

## 4. COURSE DESIGN

The developed blended learning model was used in the "C Programming" course, which is given to students in the Computer Programming program of the Afyon Kocatepe University (Turkey). The academic year in Turkey consists of an autumn and a spring term, so the course curriculum is suitable for a two-term academic year. Table 2 represents the course curriculum.

Table 2. The course curriculum for two terms

| "C Programming" Course – 1st Term | "C Programming" Course – 2nd Term |
|---|---|
| 1. Programming and C | 1. Pointers |
| 2. Fundamentals of C Programming | 2. Sorting Algorithms – 1 (Selection, Insertion) |
| 3. Data Types | 3. Sorting Algorithms – 2 (Bubble, Shell, Quick) |
| 4. Numeral Systems | 4. Searching Algorithms |
| 5. Variables and Constants | 5. Structural Data Types |
| 6. Data Type Transformations | 6. File Operators – 1 (Text) |
| 7. Operators | 7. File Operators – 2 (Binary) |
| 8. Basic Input / Output Functions | 8. Determiners |
| 9. Program Control | 9. Dynamic Memory Operations |
| 10. Loops – 1 (For) | 10. Graphic |
| 11. Loops – 2 (Do…While and While) | 11. Ports |
| 12. Preprocessors | 12. Basic Functions in C – 1 (Display, String) |
| 13. Functions | 13. Basic Functions in C – 2 (Math, Number, Date, Time) |
| 14. Arrays | 14. Basic Functions in C – 3 (Directory, Data Type Trans.) |

The number of students enrolled for the course is 60. The value is appropriate for this study and its objectives. "C Programming" is a substantially important course for computer programming students on account of learning fundamentals of computer programming, structure of C programs and writing reasonable programs by using C programming language. In the course, students must learn that C programs include some necessary elements to ensure an appropriate computer program and these elements must be taken into account during the transition from algorithm to C program codes. It is also important for students to have ability and vision to design and develop stable and perceptible program by using structures of the C programming language. The C Programming course also gives a chance to learn fundamentals of computer programming.

Students should be able to do followings after the completion of the course:

- Choosing the appropriate algorithm design and C program code structure for problems.
- Writing C program codes for basic programming problems.
- Solving programming problems by using specific C program code structures.
- Writing alternative program codes for specific problems by using flexible C programming language structure.
- Using more advanced C program code structures to solve advanced problems.

### 4.1. Learning approach

In the "C Programming" course, program-flow model approach of blended learning has been used. Program-flow model is more suitable for structure and objective of the study. At the beginning of the course, face to face education part is performed. After the completion of this part, the course lessons continue with e-learning part on the LMS system: @KU-UZEM. In face to face education part, teachers introduce the concepts that students should assimilate and

115

discuss about important parts of the given lessons. Objectives of each lesson and using features and functions of @KU-UZEM and intelligent tools are also introduced in this part.

E-learning part of the course is completed on @KU-UZEM. Students perform e-learning activities by using standard tools and intelligent learning environments integrated to the system. Students study course lessons, communicate with other classroom members and make online discussions by using standard e-learning tools. But these activities are some kind of subsidizers of the activities that are performed on intelligent learning environments. E-learning activities performed on @KU-UZEM are mostly based on intelligent learning environments. The activities that are performed via CTutor and ITest are organized and announced to students by teachers. Figure 5 represents the blended learning plan of the course.

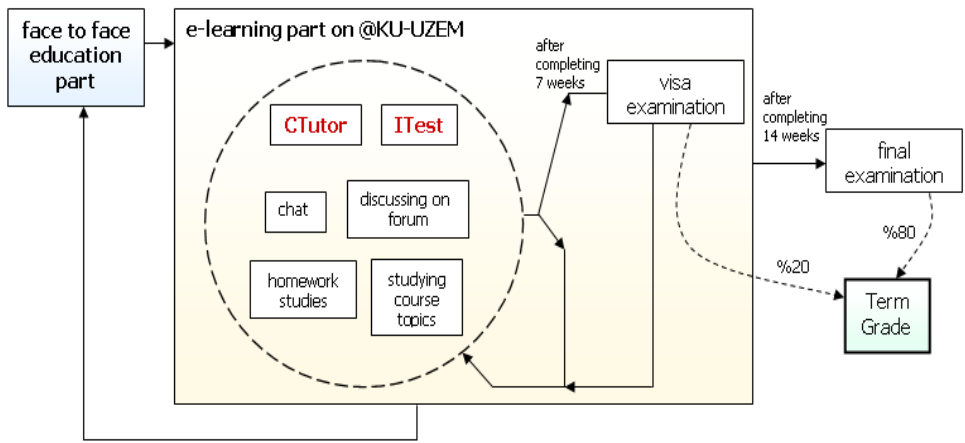

Figure 5. The blended learning plan of the course

### 4.1.1. E-Learning activities

E-learning activities performed on @KU-UZEM can be explained in more detail as below:

- Course lessons discussed in the face to face education part are also presented for students through @KU-UZEM. Online forms of lessons include some interactive elements like flash animations, animated exercises and videos that can be viewed or played by students. Students can view these lessons and continue to learning process on their own.

- Students use the chat tool integrated to @KU-UZEM to discuss about course lessons with other students and the teacher. As default, chat tool is disabled for student use and it can be enabled by only teacher. Arranged chat meetings can be announced by teachers during the face to face education part or on interface of the @KU-UZEM.

- In addition to the chat tool, the discussion forum is used by students to post new messages or make discussions on posted ones with other students and the teacher. Forum discussion topics are related to the given course lessons and online discussions are moderated by only teacher.

- For some course lessons, the teacher may give one or more homework studies to students. Homework announcements can be made during the face to face education part or via @KU-UZEM interface. Students can send prepared homework files to the teacher by using the @KU-UZEM.

International Journal of Artificial Intelligence & Applications (IJAIA), Vol.3, No.1, January 2012

- Students use the CTutor to solve C programming exercises, which are developed and provided by teachers. Each course lesson page may include one or more exercises that students can take whenever they want. It is too important for students to take all exercises that were added to the system. So, teachers make announcements for each new exercise and track students' learning activities on the CTutor system.

- By using the ITest, students take some quizzes to check their own knowledge and learning level for the given C programming course lessons. Each course lesson given to students ends with a quiz provided on ITest. Teachers can make announcements for the added quizzes by using the @KU-UZEM interface. As explained before, quiz questions are chosen automatically by the ITest. In this way, each student takes a special quiz, which was prepared according to his / her own learning level.

### 4.1.2. Grading

For each term, visa and final examinations are performed to evaluate students' success levels for the given course. The visa examination is held after completing 7 week-lesson process and students are enabled to take this examination at their home. However, final examination is held at computer laboratories after completing 14 week-lesson process. Both visa and final examinations are organized and provided on @KU-UZEM. Each examination consists of 25 multi-choice questions, which are chosen from the question bank included in the @KU-UZEM. Results of the examinations are used to determine each student's term grade. The term grade is determined by summing 20 percent of visa examination point and 80 percent of final examination point. Term grades are expressed according to the grade scale used at Afyon Kocatepe University. Students with less than 60 grade points fail and they must take the course again in the next academic year.

Students can also gain points by performing some standard e-learning activities provided on @KU-UZEM. Each activity is evaluated by the teacher in several ways and rewarded with points. At the end of the term, average points gained for each e-learning activity are calculated and added to final examination point. Table 3 shows the point values that students can gain for each activity.

Table 3. The course curriculum for two terms

| E-learning activity | Point Values to Gain | |
|---|---|---|
| | Minimum | Maximum |
| Homework | 0 | 25 |
| Forum posts | 0 | 10 |
| Participation to chat sessions | 0 | 5 |

## 5. EVALUATION

A survey was conducted at the end of the course to find out to what extent the students were accepting the blended learning model supported with @KU-UZEM and to discover students' attitude towards intelligent learning environments. Additionally, an experiment formed with an experimental group who took an active part in the realized model and a control group who only took the face to face education was performed during the first term of the given course.

117

International Journal of Artificial Intelligence & Applications (IJAIA), Vol.3, No.1, January 2012## 5.1. Student survey

Within the evaluation process, a list of 50 statements was prepared to form core of the student survey. The students were asked to express their opinion on the 1 - 5 Likert scale, checking 1 if they strongly disagree, 2 if they disagree, 3 if they have no clear opinion, 4 if they agree and 5 if they strongly agree with the given statements. The survey was anonymous and the number of respondents to the survey was 60 students (all students enrolled for the "C Programming" course). The students were asked not only to give responses for the survey statements, but also write down their comments about the blended learning model, @KU-UZEM and the intelligent learning environments. Obtained survey results have helped the authors in deciding how to focus on the future works and how to continue the development of intelligent learning environments, @KU-UZEM and the realized blended learning model.

### 5.1.1. Survey results

According to the obtained results, students accept the realized blended learning model and they are also satisfied with the intelligent learning environments and @KU-UZEM. Some statements from the survey and responses given for these statements are presented in Table 4.

Table 4. Some statements from the survey and given responses

| St. No: | Statement: | Number of responses for: | | | | | Avg. |
|---|---|---|---|---|---|---|---|
| | | 1 | 2 | 3 | 4 | 5 | |
| 3 | This learning model is more effective than traditional approaches. | 0 | 0 | 3 | 7 | 50 | 4,78 |
| 8 | The system is not good at teaching C programming language. | 47 | 7 | 5 | 1 | 0 | 1,33 |
| 17 | My academic achievement improved with this learning model. | 0 | 0 | 3 | 6 | 51 | 4,80 |
| 19 | I enjoyed the learning process on CTutor. | 0 | 1 | 3 | 6 | 50 | 4,75 |
| 24 | This learning model should be used for other programming courses. | 0 | 1 | 2 | 9 | 48 | 4,73 |
| 28 | Quizzes and exercises provided by ITest help me to learn more efficiently. | 0 | 0 | 2 | 11 | 47 | 4,75 |
| 32 | @KU-UZEM employs effective intelligent learning environments. | 0 | 0 | 5 | 10 | 45 | 4,67 |
| 37 | @KU-UZEM is an easy to use learning | 0 | 2 | 6 | 9 | 43 | 4,5 |
| | management system. | | | | | | 5 |
| 46 | I don't want to take part in this kind of study again. | 49 | 9 | 1 | 1 | 0 | 1,23 |
| 48 | I can learn faster by using the intelligent tools. | 0 | 3 | 3 | 12 | 42 | 4,55 |
| **Number of students to the survey = 60,   Avg. = Average** | | | | | | | |

According to the obtained survey results, ninety percent of students are satisfied with the learning approach implemented within this study. Eighty-three percent of students think that the @KU-UZEM provides an efficient learning platform and enables students to have better learning experiences. According to eighty-eight percent of students, CTutor is the most effective intelligent learning environment of the developed system. The survey results also show that eighty-seven percent of students are satisfied with the performed e-learning activities on @KU-UZEM. Moreover, about ninety percent of students think that the intelligent learning environments have the biggest role on effectiveness of the e-learning activities. About eighty percent of students also think that each intelligent learning environment provides a good human - computer interaction.

### 5.1.2. Student comments

Right after the survey, students wrote down their comments about the realized blended learning model, @KU-UZEM and the intelligent learning environments: CTutor and ITest. The most remarkable results that can be obtained from student comments can be listed as below:

- @KU-UZEM has a simple and colorful interface, so everyone can use it easily.
- The model causes students to get higher examination points.
- E-learning activities performed on @KU-UZEM allow students to boost their self-confidence.
- By using CTutor, students can improve their knowledge on C programming.
- CTutor provides a good learning experience for students.
- Students can check their own knowledge level and learning process with the provided questions on ITest.
- Students can use both CTutor and ITest easily.
- The realized model encourages students to study harder on course lessons.

### 5.2. Experimental evaluation

The experimental evaluation was performed during the first term of the given course. A total of 120 students participated in the experiment. Within the experiment, 60 students formed the experimental group (the group that took the course via developed blended learning model). The control group was formed with other remaining 60 students and these students only took the traditional-face to face course. In order to get accurate results, both two groups took the same visa and final examinations. Furthermore, both experimental and control group students didn't know that an experimental evaluation was performed. On the other hand, students that take part in the related groups were mostly male and they were also about 20 years old. Finally, general

119

International Journal of Artificial Intelligence & Applications (IJAIA), Vol.3, No.1, January 2012

academic achievements of each group were also at similar levels. Table 5 shows the results of both experimental and control groups.

Table 5. Comparison of the results between experimental and control groups

|  | Number of students | Students who passed (%) | Mean | Stdev. | Median |
|---|---|---|---|---|---|
| Experimental group | 60 | 82 % | 79,50 | 15,35 | 84,25 |
| Control group | 60 | 53 % | 55,69 | 14,20 | 60,05 |

As shown in Table 5, the first row represents the data of the experimental group. On the other hand, the information regarding the control group is represented in the second row. The number of students who took each test is shown in the second column whereas the percentage of the students who passed the "C Programming" course (with the term grade equal to or greater than 60) is shown in the third column. The last three columns show the mean term grade of each group, its standard deviation and median, respectively. According to the experiment results, the percentage of the students who passed the given course is significantly high in the experimental group. The mean term grade of the experimental group is also higher than the value of the control group. Eventually, experiments results point an improved student performance and knowledge level after using the intelligent learning environments and the blended learning model.

A statistical analysis has also been performed to see whether the results between the experimental group and the control group term grades were similar or not. For this purpose, the classical statistical hypothesis test (independent samples t-test) was used. The results suggest that we cannot reject the alternative hypothesis (H1), which states that the means of the two samples are different, with 95% confidence. Obtained statistical results with the independent samples test are also shown in Figure 6.

**Independent Samples Test**

| | | Levene's Test for Equality of Variances | | t-test for Equality of Means | | | | | 95% Confidence Interval of the Difference | |
|---|---|---|---|---|---|---|---|---|---|---|
| | | F | Sig. | t | df | Sig. (2-tailed) | Mean Difference | Std. Error Difference | Lower | Upper |
| Grade | Equal variances assumed | ,082 | ,775 | -8,823 | 118 | ,000 | -23,82167 | 2,70009 | -29,16858 | -18,47475 |
| | Equal variances not assumed | | | -8,823 | 117,294 | ,000 | -23,82167 | 2,70009 | -29,16892 | -18,47442 |

Figure 6. Obtained statistical results with independent samples test

Figure 7 illustrates a bar graph, which shows number of control and experimental group students achieving term grade ranges.

120

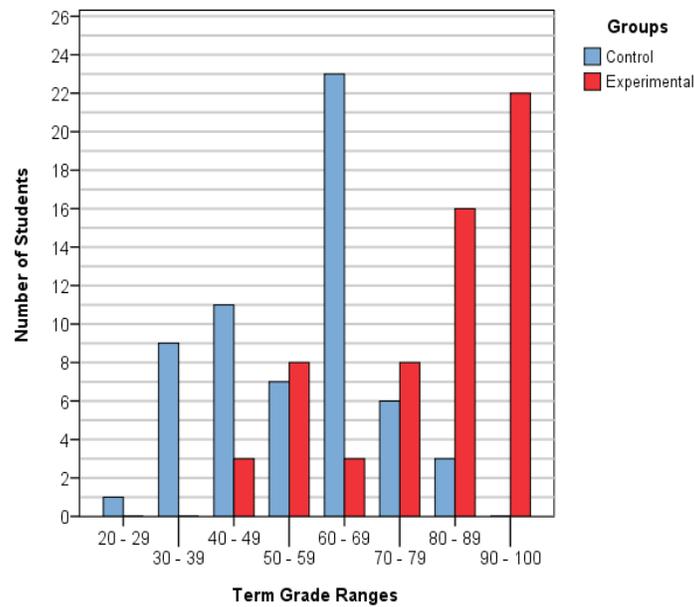

Figure 7. Number of control and experimental group students achieving term grade ranges

## 6. CONCLUSIONS

The main objective of this study was to try to improve students' success level on the "C Programming" course given at Computer Programming program of the Afyon Kocatepe University. In this aim, a blended learning model has been designed and developed to be used for the course. The model has been formed by using a combination of face to face education and e-learning parts. E-learning activities have been performed on a LMS called @KU-UZEM. In addition to standard activities performed on @KU-UZEM, the students have used two different intelligent learning environments: CTutor, a problem-solving environment; and ITest, an assessment system. At this point, the objective was to explore the contribution of two intelligent learning environments for improving students' knowledge about C programming.

According to the results obtained with the survey, students were satisfied with the blended learning approach, and their academic achievements were also better than expected. According to students' responses, used intelligent learning environments are also successful at improving students' knowledge level and academic achievements. In addition to the survey results, the performed experiment also points an improved student performance and knowledge level after using the intelligent learning environments and the blended learning model.

Regarding to the future works, the realized blended learning model will be adapted and employed in other programming courses given at the Afyon Kocatepe University. In this aim, using features and functions of the CTutor will be improved to use it in the given programming courses like "C# Programming", "Object Oriented Programming" and "Assembly". The model and the intelligent learning environments will also be adapted to some master degree programming courses given at the Institute of Science.

In addition to the development of the intelligent learning environments, features of the @KU-UZEM will also be improved to provide better e-learning experiences for students. In this aim, a web conference module is currently being developed. This module will enable teachers to conduct online meetings within the e-learning activities. The module will include live communication via web cams, whiteboard tool, text chat and desktop sharing features. Both teachers and students will be able to present slide shows or other visual works by sharing their desktop views over the web conference module. Each meeting or presentation session will be saved on @KU-UZEM database thus older sessions can be watched later.

Another work on the development of @KU-UZEM is to add mobile learning (m-learning) support to the system. Both teachers and students will be able to use their own mobile devices to perform several e-learning activities on @KU-UZEM. These activities will also include using the provided intelligent learning environments. On the other hand, RSS feed feature will be added to the system to inform teachers and students about latest announcements and updates.

There are also some works to provide a new intelligent learning environment on the @KU-UZEM. The intelligent learning environment will be a performance tracing module that uses artificial intelligence techniques to watch and evaluate students' learning activities on the @KU-UZEM. Intelligent performance tracing module will also allow watching teachers' activities to evaluate their teaching performance.

**Authors**

**Utku Kose:** Utku Kose received the B.S. degree in 2008 from computer education of Gazi University, Turkey. He received M.S. degree in 2010 from Afyon Kocatepe University, Turkey and continues Ph. D. at Selcuk University in field of computer engineering. He is currently a Lecturer in Afyon Kocatepe University. His research interest includes artificial intelligence, the chaos theory, distance education and computer education.

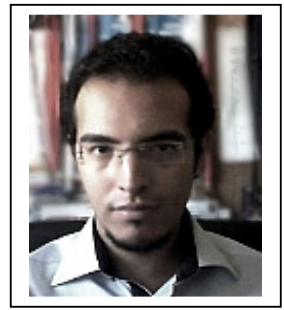

**Omer Deperlioglu:** Omer Deperlioglu received the B.S. degree in 1988 from the electrical education of Gazi University, in Turkey. He received M.S. degree in 1996 from Afyon Kocatepe University, Turkey and he completed Ph. D. degree in 2001 at Gazi University in field of controlling switch-mode dc-dc converters with neuro-fuzzy system.

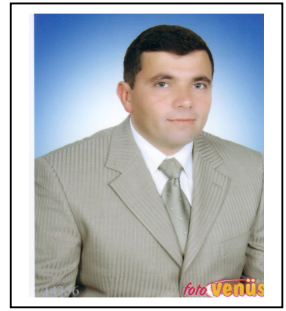

He is currently an Assistant Professor in Afyon Kocatepe University, Turkey. His research interest includes computer-based control systems, fuzzy logic control, neuro-fuzzy control and distance learning.